\documentclass[envcountsame]{llncs}
\usepackage{ae}

\usepackage{amsmath}
\usepackage{amsfonts}
\usepackage{amssymb}

\usepackage{stmaryrd}

\title{Definable Functions in the Simply Typed $\lambda$-Calculus}

\author{Mateusz Zakrzewski}
\institute{Institute of Informatics,\\Warsaw University,\\Banacha 2, 02-097 Warsaw, Poland.\\
\email{mz201114@students.mimuw.edu.pl}
}

\newcommand{\ifte}[3]{\mathtt{if}\ #1\ \mathtt{then}\ #2\ \mathtt{else}\ #3}
\newcommand{\ifzero}[3]{\mathtt{ifzero}\ #1\ \mathtt{then}\ #2\ \mathtt{else}\ #3}

\newcommand{\numbers}{n_{1},n_{2},\ldots,n_{k}}
\newcommand{\numerals}{\rho(n_{1})\rho(n_{2})\ldots\rho(n_{k})}

\newcommand{\inmodel}[1]{\llbracket #1\rrbracket^{\mathcal{M}}}

\begin{document}

\maketitle

\begin{abstract}
It is a common knowledge that the integer functions definable in the simply typed $\lambda$-calculus are exactly the extended polynomials. This is indeed the case when one interprets integers over the type $(p\to p)\to p\to p$ where $p$ is a base type and/or equality is taken as \linebreak $\beta$-conversion. It is commonly believed that the same holds for\linebreak $\beta\eta$-equality and for integers represented over any fixed type of the form $(\tau\to \tau)\to \tau\to \tau$. 

In this paper we show that this opinion is not quite true. We prove that the class of functions strictly definable in the simply typed $\lambda$-calculus is considerably larger than the extended polynomials.
Next, we investigate which functions belong to that class.

\end{abstract}

\section{Introduction}
We assume that all types are constructed from only one base type, which we denote by $o$. Natural numbers are represented as Church numerals: $0$=$\lambda fx.x$, $1$=$\lambda fx.fx$, $2$=$\lambda fx.f(fx)$, $3$=$\lambda fx.f(f(fx))$, \ldots Every type that can be assigned to all Church numerals is of the form $(\tau\to\tau)\to\tau\to\tau$, for some $\tau$.

We say that a function $f$ over a free algebra $A$, of arity $k$, is \emph{definable} (or~\emph{representable}) in the untyped $\lambda$-calculus, if there exists a term $E$, such that $E\rho(a_{1})\rho(a_{2})\ldots\rho(a_{k})$ is equal (up to $\beta$- or $\beta\eta$-conversion) to $\rho(f(a_{1},a_{2},\ldots,a_{k}))$ for all $a_{1},a_{2},\ldots,a_{k}\in A$. By $\rho(a)$ we mean the term which represents $a$ (as~defined in~\cite{leiv}) e.g. if $A=\mathbb{N}$ and $a\in A$ then $\rho(A)$ is a Church numeral.
In typed $\lambda$-calculi, there are additional requirements that must be met by a definable function. We~say that a function is \emph{non-uniformly} definable, if $E\rho(a_{1})\rho(a_{2})\ldots\rho(a_{k})$ can be typed for all $a_{1},a_{2},\ldots,a_{k}$. A function is \emph{skewly} definable, if there are types $\tau_{1},\tau_{2},\ldots,\tau_{k}$, such that $E\rho(a_{1})\rho(a_{2})\ldots\rho(a_{k})$ can be typed for all $a_{1},a_{2},\linebreak\ldots,a_{k}$, with type $\tau_{i}$ assigned to $\rho(a_{i})$ ($i=1\ldots k$). If $\tau_{i}$ are all the same and equal to the type of $E\rho(a_{1})\rho(a_{2})\ldots \rho(a_{k})$, we say that $E$ represents $f$ \emph{strictly} (or $f$ is strictly definable).

In the simply typed $\lambda$-calculus, functions over natural numbers that are strictly definable with $\tau$ being a base type (or a type variable) have been characterized by Schwichtenberg (see~\cite{schw}) and are known as the extended polynomials. The result of Schwichtenberg has been generalized to arbitrary free algebras~\cite{zai}. Leivant has shown that, in simple types, every non-uniformly definable function (over any algebra) is skewly definable. The exact characterization of skewly definable functions is not known. It has been shown (see~\cite{leiv}) that when equality is taken as $\beta$-conversion, every function that is strictly definable is strictly definable with $\tau$ being a base type. It was also claimed in~\cite{leiv} that the same holds when equality is taken as $\beta\eta$-conversion. Surprisingly, as we show below, it is not true even for functions over natural numbers. In other words, there exist strictly definable functions that are not extended polynomials.

This paper aims to investigate which functions are strictly definable over natural numbers, with arbitrary $\tau$ and $\beta\eta$-conversion.
We define $\mathcal{F}$ as the class of strictly definable functions and $\mathcal{G}$ as a class that contains extended polynomials and two additional functions (or, more precisely, two function schemas) and is closed under composition. We prove that for every function $f\in\mathcal{G}$, there exists a type $\tau$ such that $f$ is definable with numerals of type $(\tau\to\tau)\to\tau\to\tau$. It follows that $\mathcal{G}$ is a subset of $\mathcal{F}$. The proof yields an effective procedure that can be used to find $\tau$ and a term which represents $f$. 

We conjecture that $\mathcal{G}$ exactly characterizes strictly definable functions, i.e. $\mathcal{G}=\mathcal{F}$, and we gather some evidence for this conjecture proving, for example, that every skewly representable finite range function is strictly representable over $(\tau\to \tau)\to \tau\to \tau$, for some~$\tau$. We also prove that all definable functions of the form: $f(m,n_{1},n_{2})=\ifte{m\in A}{m_{1}}{m_{2}}$ (for all $A\subseteq \mathbb{N}$) are in $\mathcal{G}$. Finally, we give examples of very simple functions that are not definable.

\subsection{Extended Polynomials}

\begin{definition}\label{ext_pol_def}\rm
The class of \emph{extended polynomials} is the smallest class of functions over $\mathbb{N}$ which contains:
\begin{enumerate}
\item the constant functions: $0$ and $1$,
\item projections,
\item addition,
\item multiplication,
\item the function $\mathrm{ifzero}(n,m,p)=\ifte{n=0}{m}{p}$,
\end{enumerate}
and is closed under composition.
\end{definition}

Addition, multiplication and ifzero can be represented by the terms:
\linebreak
$\lambda nm.\lambda fx.nf(mfx)$, $\lambda nm.\lambda fx.n(mf)x$ and $\lambda nmp.\lambda fx.n(\lambda y.pfx)(mfx)$, respectively. For all $\tau$, every $k$-ary extended polynomial can be strictly represented by a term of type $\omega_{\tau}^{k}\to\omega_{\tau}$ where $\omega_{\tau}=(\tau\to\tau)\to\tau\to\tau$.

\subsection{Beyond Extended Polynomials}
Consider the following term:\\

\pagebreak
\indent
$E=\lambda n.\lambda fx.$ \\
\indent\indent
$\lambda a_{1}a_{2}\ldots a_{l}.$\\
\indent\indent\indent
$(n\;(\lambda y.\lambda z_{1}z_{2}\ldots z_{l}.yz_{2}z_{3}\ldots z_{l}z_{1})\;(\lambda z_{1}z_{2}\ldots z_{l}.z_{1}))
$\\
\indent\indent\indent\indent
$(x\ a_{1}a_{2}\ldots a_{l})$\\
\indent\indent\indent\indent
$(fx\ a_{1}a_{2}\ldots a_{l})$\\
\indent\indent\indent\indent
$(f(fx)\ a_{1}a_{2}\ldots a_{l})$\\
\indent\indent\indent\indent
$\ldots$\\
\indent\indent\indent\indent
$((f^{l-1}x)\ a_{1}a_{2}\ldots a_{l})$.\\\\

\noindent If $n$ is assigned a numeral $N$, the term $$(n\;(\lambda y.\lambda z_{1}z_{2}\ldots z_{l}.yz_{2}z_{3}\ldots z_{l}z_{1})
\;(\lambda z_{1}z_{2}\ldots z_{l}.z_{1}))$$ 

\noindent will evaluate (or reduce) to
$(\lambda z_{1}z_{2}\ldots z_{l}.z_{i})$, where $i$ is equal to $(N\bmod{l})+1$. Therefore, $E$ strictly represents the function $g(n)=n\bmod{l}$ and can be assigned the type $\omega_{\tau}\to\omega_{\tau}$, where $\tau$ must be of the form $\alpha^{l}\to\alpha$, for some $\alpha$. 

\begin{proposition}\label{pol_inc}
For every extended polynomial $f:\mathbb{N}\to\mathbb{N}$, there exists an $m$, such that for all $n_{1},n_{2}\geq m$, if $n_{1}\geq n_{2}$ then $f(n_{1})\geq f(n_{2})$.
\end{proposition}

\begin{proof}
Simple induction with respect to the structure of a polynomial.
\end{proof}

Proposition~\ref{pol_inc} states that every extended polynomial is non-decreasing for sufficiently large arguments. Clearly, it is not the case for $g$. It follows that $g$ is not an extended polynomial. It is worth noting that for this example to work, we need both $\beta\eta$-conversion and numerals of type $\omega_{\tau}$ with $\tau$ being different from the base type.

\section{Strictly Definable Functions}

In this section, we define a class of functions and prove that all functions which are in that class are (strictly) definable.

\begin{definition}\label{g_def}\rm
$\mathcal{G}$ is the smallest class of functions which is closed under composition and contains:

\begin{enumerate}
\item extended polynomials,
\item the function $f_{1}^{l}(m,n_{1},n_{2},\ldots,n_{l})=n_{i}$, where 
$i=(m\bmod{l})+1$,\\(for all $l\geq 2$),
\item the function $f_{2}^{l}(m,n_{1},n_{2})=\mathtt{if}\ m\leq l\ \mathtt{then}\ n_{1}\ \mathtt{else}\ n_{2}$, (for all $l\geq 1$).
\end{enumerate}

\end{definition}

By $\mathcal{F}$ we denote the class of all functions that are strictly definable. Notation $(s)\langle M_{1},M_{2},\ldots,M_{s}\rangle$ means a tuple of length $s$, represented by the term $\lambda p.pM_{1}M_{2}\ldots M_{s}$. By $\Pi_{i}(s)(p)$, we denote the $i$-th element of the tuple $p$ of length $s$, represented by $p(\lambda x_{1}x_{2}\ldots x_{s}.x_{i})$. The function $f_{2}^{l}$ can be represented by the term:

$$E=\lambda mn_{1}n_{2}.\lambda fx.\lambda a_{1}a_{2}a_{3}.\ifzero{Gm}{n_{1}fxa_{1}a_{2}a_{3}}{n_{2}fxa_{1}a_{2}a_{3}},$$

\noindent where:

$\ifzero{x}{y}{z}=x(\lambda t.z)y$,

$G=\lambda m.\Pi_{1}(l+1)(mFP)$,

$F=\lambda p.\langle \Pi_{2}(l+1)(p),\Pi_{3}(l+1)(p),\ldots,\Pi_{l+1}(l+1)(p),\mathtt{succ}\ \Pi_{l+1}(l+1)(p)\rangle$,

$P=(l+1)\langle 0,0,\ldots,0\rangle$,

$\mathtt{succ}=\lambda n.\lambda fx.f(nfx)$.\\

\noindent The term $G$ skewly represents the function $g(m)=m-l$. 
Therefore $\ifzero{\linebreak Gm}{x}{y}$
returns the same value as $\ifte{m\leq l}{x}{y}$. \\

\noindent Here is how $E$ can be typed:\\

\indent \indent $\vdash E:\omega_{\tau}^3\to\omega_{\tau}$,

\indent \indent $\tau=(\alpha^{l+1}\to\alpha)\to\alpha$,

\indent \indent $\alpha=\omega_{\tau^{\prime}}$.\\

The type $\tau$ is the type of an $(l+1)$-tuple of numerals (hence $\alpha=\omega_{\tau^{\prime}}$, for some $\tau^{\prime}$). To prove that all functions which are in $\mathcal{G}$ are definable, we will need more flexibility as to what types can be assigned to the terms which represent them. For example, if (for \emph{any} $s\geq l+1$) the function $g$ is represented like this:\\

\indent\indent$G=\lambda n.\Pi_{s-l}(s)(nFP)$,

\indent\indent$F=\lambda p.(s)\langle \Pi_{2}(s)(p),\Pi_{3}(s)(p),\ldots,\Pi_{s}(s)(p),\mathtt{succ}\ \Pi_{s}(s)(p)\rangle$,

\indent\indent$P=(s)\langle 0,0,\ldots,0\rangle$.\\

\noindent then $E$ can be assigned the type $\omega_{\tau}^3\to\omega_{\tau}$ with $\tau$ being the type of a tuple of length $s$. A flexible representation of $f_{1}^{l}$ is as follows:

$$E=\lambda m.\lambda n_{1}n_{2}\ldots n_{l}.\lambda fx.
\lambda a.\Pi_{l}(s)(mFP),$$

\noindent where:

\indent $F=\lambda p.(s)\langle \Pi_{2}(s)(p),\Pi_{3}(s)(p),\ldots,\Pi_{l}(s)(p),\Pi_{1}(s)(p),\Pi_{l+1}(s)(p)$

\indent \indent \indent \indent
$\,\,\,\,\Pi_{l+2}(s)(p),\ldots,\Pi_{s}(s)(p)\rangle$,

\indent $P=(s)\langle M_{2},M_{3},\ldots,M_{l},M_{1},X,X,\ldots,X\rangle$,

\indent $X=xa$,

\indent $\forall_{1\leq i\leq l}\;\;M_{i}=n_{i}fxa$.\\

The term $E$, when applied to arguments, evaluates similarly to the term which represents the remainder modulo $l$ (as given in the introduction). The main difference is that we can now use tuples of arbitrary length $s\geq l$. We can assign the type $\omega_{\tau}^{l+1}\to \omega_{\tau}$ (with $\tau=(\alpha^{s}\to \alpha)\to \alpha$) to $E$.

It is easy to see that $\mathcal{G}$ can be equivalently defined as follows:

\begin{definition}\label{2_g_def}\rm
$\mathcal{G}$ is the smallest class of functions which is closed under composition and contains:

\begin{enumerate}
\item the constant functions: $0$ and $1$,
\item projections,
\item addition,
\item multiplication,
\item the function $\mathrm{ifzero}(n,m,p)=\ifte{n=0}{m}{p}$,
\item the function $f_{1}^{l}(m,n_{1},n_{2},\ldots,n_{l})=n_{i}$, where 
$i=(m\bmod{l})+1$, (for all $l\geq 2$),
\item the function $f_{2}^{l}(m,n_{1},n_{2})=\mathtt{if}\ m\leq l\ \mathtt{then}\ n_{1}\ \mathtt{else}\ n_{2}$, (for all $l\geq 1$).
\end{enumerate}

\end{definition}

\begin{definition}\label{3_g_def}\rm
$\mathcal{G}$ is the smallest class of functions such that for all $k$:
\begin{enumerate}
\item if $\;\forall\vec{n}\;f(\vec{n})=0$ then $f\in \mathcal{G}$,
\item if $\;\forall\vec{n}\;f(\vec{n})=1$ then $f\in \mathcal{G}$,
\item $\forall_{1\leq i\leq k}\;\;$if $\;\forall\vec{n}\;f(\vec{n})=n_{i}$ then $f\in \mathcal{G}$,
\item if $g_{1},g_{2}\in \mathcal{G}\;$ and  $\;f(\vec{n})=g_{1}(\vec{n})+g_{2}(\vec{n})$, then $f\in \mathcal{G}$,
\item if $g_{1},g_{2}\in \mathcal{G}\;$ and  $\;f(\vec{n})=g_{1}(\vec{n})\cdot g_{2}(\vec{n})$, then $f\in \mathcal{G}$,
\item if $g,h_{1},h_{2}\in \mathcal{G}\;$ and 
$\;f(\vec{n})=\ifte{g(\vec{n})=0}{h_{1}(\vec{n})}{h_{2}(\vec{n})}$, then $f\in \mathcal{G}$,
\item $\forall_{2\leq l\leq k-1}\;\;$if $g\in \mathcal{G}$, $\forall_{1\leq i\leq l}\;\;h_{i}\in \mathcal{G}\;$ and $\;f(\vec{n})=h_{j}(\vec{n})$, where $j=(g(\vec{n})\bmod{l})+1$, then $f\in \mathcal{G}$,
\item $\forall_{l\geq 1}\;\;$if $g,h_{1},h_{2}\in \mathcal{G}\;$ and 
$\;f(\vec{n})=\ifte{g(\vec{n})\leq l}{h_{1}(\vec{n})}{h_{2}(\vec{n})}$, then $f\in \mathcal{G}$,
\end{enumerate}
where $\vec{n}=n_{1},n_{2},\ldots,n_{k}$.
\end{definition}

\begin{lemma}\label{equiv_def}
Definition~\ref{3_g_def} is yet another equivalent definition of $\mathcal{G}$:
\end{lemma}

\begin{proof}
Let $\mathcal{G}_{1}$ be the class specified by Definition~\ref{2_g_def} and let $G_{2}$ be the class specified by Definition~\ref{3_g_def}. A simple induction on Definition~\ref{2_g_def} is enough to prove that $\mathcal{G}_{1}\subseteq \mathcal{G}_{2}$. It is easy to see that $\mathcal{G}_{2}\subseteq \mathcal{G}_{1}$.
\end{proof}

\begin{theorem}\label{composition}
For all $k$-ary functions $f\in\mathcal{G}$ and for almost all $s$, there exists an~$E$ such that:
\begin{enumerate}
\item $\forall \numbers\;\; E\numerals=_{\beta\eta}\rho(f(\numbers))$,
\item $\vdash E:{\omega_{\tau(s)}}^{k}\to \omega_{\tau(s)}$,
\end{enumerate}
where $\tau(s)=(\alpha^{s}\to\alpha)\to\alpha$ with $\alpha=\omega_{o}$ (i.e. $\tau(s)$ is the type of an $s$-tuple of numerals). 
\end{theorem}

\begin{proof}
Induction on Definition~\ref{3_g_def}. Here we thoroughly discuss one case (the other cases are similar). Suppose $f(\vec{n})=g_{1}(\vec{n})+g_{2}(\vec{n})$, where $g_{1},g_{2}\in \mathcal{G}$. By the induction hypothesis:

\pagebreak
\noindent For almost all $s$, there exists a $G_{1}$ such that:
\begin{enumerate}
\item $\forall \numbers\;\; G_{1}\numerals=_{\beta\eta}\rho(g_{1}(\numbers))$, 
\item $\vdash G_{1}:{\omega_{\tau(s)}}^{k}\to \omega_{\tau(s)}$.
\end{enumerate}

\noindent For almost all $s$, there exists a $G_{2}$ such that:
\begin{enumerate}
\item $\forall \numbers\;\; G_{2}\numerals=_{\beta\eta}\rho(g_{2}(\numbers))$,
\item $\vdash G_{2}:{\omega_{\tau(s)}}^{k}\to \omega_{\tau(s)}$.
\end{enumerate}

It follows that for almost all $s$, 
the function $f$ can be represented by the term $E=\lambda \vec{n}.\lambda fx.(G_{1}\vec{n})f(G_{2}\vec{n} fx)$, which is similar to the term $\lambda n_{1}n_{2}.\lambda fx.n_{1}f(n_{2}fx)$ (the one used to represent addition), but there are two differences:
\begin{enumerate}
\item it takes $k$ arguments of type $\omega_{\tau(s)}$ instead of two,
\item $n_{i}$ is replaced with $G_{i}\vec{n}$ (i=1,2) in the part to the right of ``$\lambda fx$''.
\end{enumerate}

The terms needed in other cases can be obtained similarly. There is always only a finite number of tuple lengths for which $E$ cannot be constructed. \qed
\end{proof}

\begin{corollary}\label{comp_cor}
$\mathcal{G}\subseteq\mathcal{F}$.
\end{corollary}

\section{Limitations of Definability}
In this section we show some properties of the class $\mathcal{G}$ to narrow the gap between the functions that are known to be definable and the functions that are known to be undefinable. We make extensive use of the following theorem (proved in~\cite{stat}).

\begin{theorem}\label{Statman}
For arbitrary $M$ of type $\alpha$, there exists a finite model $\mathcal{M}$ such that for every term $N$ of type $\alpha$:
\begin{center}
$M=_{\beta\eta}N$ if and only if $\mathcal{M}\models M=N.$
\end{center}
\end{theorem}

\begin{lemma}\label{cond_stat}
Let $\mathcal{M}$ be a finite model, and let $\alpha=\omega_{\tau}$, for some $\tau$. For every $c\in\mathcal{M}_{\omega_{\tau}}$ the function $h(m,n_{1},n_{2})=\ifte{(\inmodel{\rho(m)^{\alpha}}=c)}{n_{1}}{n_{2}}$ is definable.
\end{lemma}

\begin{proof}
Let $l$ be the smallest number such that $\inmodel{\rho(l+t)^{\alpha}}=\inmodel{\rho(l)^{\alpha}}$, for some $t\geq 1$. Let $t_{\mathrm{min}}$ be the smallest such $t$. If $c=\inmodel{\rho(s)^{\alpha}}$, where $s<l$, then\\
 
\indent \indent $h(m,n_{1},n_{2})=$

\indent \indent $\ifte{(m=s)}{n_{1}}{n_{2}}=$

\indent \indent $\ifte{m\leq s}{(\ifte{m\leq s-1}{n_2}{n_1})}{n_{2}}$.\\

\noindent The function $h$ is a composition of functions which belong to $\mathcal{G}$. Therefore $h$~is in~$\mathcal{G}$. If $c=\inmodel{\rho(s)^{\alpha}}$, where $s\geq l$, then\\

\indent \indent $h(m,n_{1},n_{2})=$

\indent \indent $\ifte{m\equiv s \pmod{t_{\mathrm{min}}}}{n_{1}}{n_{2}}$.\\

\noindent Again, $h$ can be expressed as a composition of functions which belong to $\mathcal{G}$.\linebreak If $c\neq\inmodel{\rho(s)^{\alpha}}$ for all $s\in \mathbb{N}$ then $h(m,n_{1},n_{2})=n_{2}$. Therefore $h\in \mathcal{G}$.

\end{proof}

\begin{theorem}\label{cond_def}
If $f$ is of the form $f(m,n_{1},n_{2})=\ifte{m\in A}{n_{1}}{n_{2}}$, for some $A\subseteq \mathbb{N}$, then $f$ is not definable or $f\in\mathcal{G}$.
\end{theorem}

\begin{proof}
Suppose $f$ is definable. Therefore, $g(m)=f(m,1,0)$ is also definable.
It is clear that the following holds:
$$f(m,n_{1},n_{2})=\ifzero{g(m)}{n_{2}}{n_{1}}.$$

\noindent Let $G$ of type $\alpha\to\alpha$ be a term which represents $g$. The following is the result of applying Theorem~\ref{Statman} to $\rho(0)^{\alpha}$:
\begin{enumerate}
\item $N=_{\beta\eta}\rho(0)^{\alpha}$ if and only if $\mathcal{M}\models N=\rho(0)$ for all $N$ of type $\alpha$,
\item $\mathcal{M}$ is finite.
\end{enumerate}

\noindent Let $B=\{\inmodel{\rho(m)}\;|\;m \in \mathbb{N},\;\inmodel{G}\cdot \inmodel{\rho(m)}=\inmodel{\rho(0)^{\alpha}}\}$.
Then $B=\{b_{1},b_{2},...,b_{k}\}$, for some $k\geq 0$.\\

\noindent The~function $f$ can be expressed as follows:\\

\indent \indent $f(m,n_{1},n_{2})=$

\indent \indent\indent $\ifte{(\inmodel{\rho(m)^{\alpha}}=b_{1})}{n_{2}}{\ }$

\indent \indent\indent $\ifte{(\inmodel{\rho(m)^{\alpha}}=b_{2})}{n_{2}}{\ }$

\indent \indent\indent $...$

\indent \indent\indent $\ifte{(\inmodel{\rho(m)^{\alpha}}=b_{k})}{n_{2}}{n_{1}}$.\\

\noindent It follows from Lemma~\ref{cond_stat} that $f$ is a composition of functions which belong to~$\mathcal{G}$. Therefore $f\in \mathcal{G}$. \qed

\end{proof}

\noindent The next theorem is a generalization of Theorem~\ref{cond_def}.

\begin{theorem}\label{skew_def}
Every skewly representable finite range function is strictly representable.
\end{theorem}

\begin{proof}
Let $f$ be a $k$-ary skewly representable finite range function, and let $A$ be the range of $f$. The~function $f$ is equal to the function $f^{\prime}$, which is defined as follows:\\

\indent \indent $f^{\prime}(\vec{m})=$

\indent \indent\indent $\ifte{(f(\vec{m})=a_{1})}{a_{1}}{\ }$

\indent \indent\indent $\ifte{(f(\vec{m})=a_{2})}{a_{2}}{\ }$

\indent \indent\indent $...$

\indent \indent\indent $\ifte{(f(\vec{m})=a_{p-1})}{a_{p-1}}{a_{p}}$,\\

\noindent where 

$A=\{a_{1},a_{2},...,a_{p}\}$,

$\vec{m}=m_{1},m_{2},\ldots,m_{k}$.\\

\noindent The function $f$ is a composition of functions of the form:
$$h_{i}(\vec{m},n_{1},n_{2})=\ifte{f(\vec{m})=a_{i}}{n_{1}}{n_{2}}.$$

We show that $h_{i}\in\mathcal{G}$, for arbitrary $1\leq i\leq p$. Let $E$ of type $\alpha_{1}\to\alpha_{2}\to\cdots\to\alpha_{k}\to\beta$ be a term which skewly represents $f$. The following is the result of applying Theorem~\ref{Statman} to $\rho(a_{i})^{\beta}$:
\begin{enumerate}
\item $N=_{\beta\eta}\rho(a_{i})^{\beta}$ if and only if $\mathcal{M}\models N=\rho(a_{i})$ for all $N$ of type $\beta$,
\item $\mathcal{M}$ is finite.
\end{enumerate}

\noindent Let $B=\{
\langle\inmodel{\rho(m_{1})^{\alpha_{1}}},
\inmodel{\rho(m_{2})^{\alpha_{2}}},
\ldots,
\inmodel{\rho(m_{k})^{\alpha_{k}}}\rangle
\;|\;m_{1},m_{2},\ldots,m_{k}\in\mathbb{N},\\
\indent\hspace{30pt}\inmodel{E}
\cdot
\inmodel{\rho(m_{1})^{\alpha_{1}}}
\cdot
\inmodel{\rho(m_{2})^{\alpha_{2}}}
\cdot
\cdots
\cdot
\inmodel{\rho(m_{k})^{\alpha_{k}}}
=
\inmodel{\rho(a_{i})^{\beta}}
\}.$\\

\noindent The set $B$ consists of the arguments at which the value of $f$ is equal to $a_{i}$, in model $\mathcal{M}$.
The~function $h_{i}$ can be expressed as follows:\\
$$h_{i}(\vec{m},n_{1},n_{2})=\ifte{\inmodel{\rho(\vec{m})}\in B}{n_{1}}{n_{2}},$$

\noindent where

$\vec{m}=m_{1},m_{2},\ldots,m_{k}$,

$\inmodel{\rho(\vec{m})}=\langle\inmodel{\rho(m_{1})^{\alpha_{1}}},
\inmodel{\rho(m_{2})^{\alpha_{2}}},
\ldots,
\inmodel{\rho(m_{k})^{\alpha_{k}}}\rangle.$\\

\noindent The set $B$ is finite. It follows from Lemma~\ref{cond_stat} that $h_{i}$ is a composition of functions which belong to~$\mathcal{G}$. Therefore $h_{i}\in \mathcal{G}$. \qed
\end{proof}

\begin{proposition}\label{pred_undef}
The predecessor function is not strictly definable.
\end{proposition}
\begin{proof}
Suppose $\mathrm{pred}(n)=n-1$ is strictly definable. Let $E$ be a term that represents $\mathrm{pred}$. Therefore, subtraction can be skewly represented by the term $\lambda nm.mEn$. On the other hand, we know that subtraction is not skewly definable (see~\cite{flo}).
\end{proof}

\begin{proposition}\label{div_undef}
Division by two is not strictly definable.
\end{proposition}
\begin{proof}
Suppose $f(n)=n\ \mathtt{div}\ 2$ is strictly definable. Let $E$ be a term that strictly represents $f$. Let $g$ be the function which is skewly represented by the term $G=\lambda nm.mEn$ of type $\alpha\to\beta\to\alpha$, where $\beta=\omega_{\alpha}$. The following is the result of applying Theorem~\ref{Statman} to $\rho(0)^{\alpha}$:
\begin{enumerate}
\item $N=_{\beta\eta}\rho(0)^{\alpha}$ if and only if $\mathcal{M}\models N=\rho(0)$ for all $N$ of type $\alpha$,
\item $\mathcal{M}$ is finite.
\end{enumerate}

Let $l$ be the smallest number such that $\inmodel{\rho(l+t)^{\alpha}}=\inmodel{\rho(l)^{\alpha}}$, for some $t\geq 1$. Let $t_{\mathrm{min}}$ be the smallest such $t$. Similarly,
let $l^{\prime}$ be the smallest number such that $\inmodel{\rho(l^{\prime}+t^{\prime})^{\beta}}=\inmodel{\rho(l^{\prime})^{\beta}}$, for some $t^{\prime}\geq 1$, and let ${t_{\mathrm{min}}^{\prime}}$ be the smallest such $t^{\prime}$. 

\pagebreak
\noindent On one hand, we have:\\

\indent\indent$\inmodel{G}\cdot\inmodel{\rho(l)^{\alpha}}\cdot\inmodel{\rho(l^{\prime})^{\beta}}=
\inmodel{G}\cdot
\inmodel{\rho(l)^{\alpha}}\cdot
\inmodel{\rho(l^{\prime}+l\cdot {t_{\mathrm{min}}^{\prime}})^{\beta}}=$

\indent\indent$\inmodel{\rho(g(
l,
l^{\prime}+l\cdot {t_{\mathrm{min}}^{\prime}}
))^{\alpha}}=
\inmodel{\rho(0)^{\alpha}}
$.\\

\noindent On the other hand, we have:\\

\indent\indent$\inmodel{G}\cdot\inmodel{\rho(l)^{\alpha}}\cdot\inmodel{\rho(l^{\prime})^{\beta}}=
\inmodel{G}\cdot
\inmodel{\rho(l+2^{l^{\prime}}\cdot t_{\min})^{\alpha}}\cdot
\inmodel{\rho(l^{\prime})^{\beta}}=$

\indent\indent$\inmodel{\rho(g(
l+2^{l^{\prime}}\cdot t_{\min},
l^{\prime}
))^{\alpha}}\neq
\inmodel{\rho(0)^{\alpha}}
$.
\end{proof}

\section{Conclusion}
We have given examples of functions that are strictly definable and are not extended polynomials. It is clear that there are many considerably different kinds of strict definability, depending on the type of numerals. There is also an unexpected difference between $\beta$- and $\beta\eta$-conversion regarding $\lambda$-definability. A~corollary of Proposition~\ref{pred_undef} is that functions which are strictly definable with arbitrary type of numerals form a proper subset of the set of skewly definable functions. Theorem~\ref{skew_def} states that it is not the case for functions with finite range.

We conjecture that $\mathcal{G}$ exactly characterizes strictly definable functions (or $\mathcal{G}=\mathcal{F}$). We have shown that $\mathcal{G}\subseteq\mathcal{F}$. The functions: $f_{1}^{l}$ and $f_{2}^{l}$ are compositions of functions of the form $\ifte{m\in A}{n_{1}}{n_{2}}$, for all $l$. It follows from Theorem~\ref{cond_def} that every strictly definable function of that form is already in $\mathcal{G}$.

\end{document}